# Electronic orbital angular momentum and magnetism of graphene

Ji Luo

*Department of Physics and Institute for Functional Nanomaterials,*
*University of Puerto Rico at Mayagüez, Mayagüez, PR 00681, USA*
E-mail: ji.luo@upr.edu

ABSTRACT
Orbital angular momentum (OAM) of graphene electrons in a perpendicular magnetic field is calculated and corresponding magnetic moment is used to investigate the magnetism of perfect graphene. Variation in magnetization demonstrates its decrease with carrier-doping, plateaus in a large field, and de Haas-van Alphen oscillation. Regulation of graphene's magnetism by a parallel electric field is presented. The OAM originates from atomic-scale electronic motion in graphene lattice, and vector hopping interaction between carbon atomic orbitals is the building element. A comparison between OAM of graphene electrons, OAM of Dirac fermions, and total angular momentum of the latter demonstrates their different roles in graphene's magnetism. Applicability and relation to experiments of the results are discussed.

## 1. Introduction

The magnetism of graphene is rather intriguing. It is believed that carrier-undoped graphene has a very large diamagnetic susceptibility and the susceptibility decreases rapidly with the increasing carrier-doping of either electrons or holes [1-3]. Theoretically, the susceptibility was derived according to quasi-continuous Landau levels (LLs) of graphene in a weak magnetic field, and the thermal potential energy constructed from LLs plays the central role [3,4]. This anomalous susceptibility is interesting and attempts have been made to interpret its origin [3].

In addition, many factors have been found to contribute to graphene's magnetism, such as vacancies [5-8], substituting atoms [8], absorbed atoms [8-10], edge structures and edge states [11-13], finite size of graphene [14,15], electron-electron interaction [2], substrates [16], and strain [17]. Magnetism-related properties of gapped graphene were also studied [18,19]. Although diamagnetism was observed in experiments, the magnetization was measured for graphene crystallites with nanometer size [20]. For a new material, the elimination of uncontrollable disturbances is necessary for both applications and the revelation of underlying physics, and graphene is expected to be made more and more perfect. Theoretically, wave functions of graphene electrons in a perpendicular magnetic field are obtainable [21,22]. Along with LLs, these wave functions may provide more insight into graphene's magnetic properties.

In this work, orbital angular momentum (OAM) of graphene electrons in a perpendicular magnetic field is calculated according to the electronic wave functions. The corresponding orbital magnetic moment (OMM) is found paramagnetic for negative LLs and diamagnetic for positive LLs. As a result, carrier-undoped graphene could be paramagnetic and the magnetization decreases with the increasing carrier-doping of either electrons or holes. For a fixed magnetic field, the magnetization variation with the carrier-doping is obtained as a function of temperature and Fermi energy. It presents plateaus similar to those in quantum Hall effect as Fermi energy varies in a large magnetic field. As the magnetic field varies, the magnetization demonstrates de Haas-van Alphen oscillation. A parallel electric field can change the electronic states and regulate graphene's magnetism. The derivation of OAM of graphene electrons manifests that their OMM originates from their atomic-scale motion in graphene lattice. The vector hopping interaction of carbon atomic orbitals constitutes building element of the OAM. Its honeycomb-like distribution in graphene lattice and its magnitude result in the unique form of the OAM operator, and its modulation by the two-component electronic wave function generates the OAM for a specific state. The OAM of a graphene electron is different from that of the Dirac fermion, although they are described by the same two-component wave function. The superposition of degenerate states may result in diversity of graphene's magnetization. This superposition, the deep states with energies far below the zero, and the small size of graphene crystallites may be the origin of experimentally observed diamagnetism.

**2. Orbital angular momentum, orbital magnetic moment, and magnetization variation**

The graphene is taken as $xy$-plane with $x$-axis parallel to one set of C-C bonds. Unit vectors of the axes are denoted by $\vec{x}$, $\vec{y}$, and $\vec{z}$. A three-dimensional vector is denoted by $\vec{r} = x\vec{x} + y\vec{y} + z\vec{z}$ and its two-dimensional projection in the graphene plane with $z=0$ by $\vec{\rho} = x\vec{x} + y\vec{y}$. Two kinds of Dirac points $\vec{k}_F = (2\pi/3\sqrt{3}a_0)(\sqrt{3}\vec{x} + \tau\vec{y})$ are distinguished by $\tau = \pm 1$, with $a_0$ the C-C bond length. A graphene electron is described by a two-component wave function $\Psi = (\psi_1\ \psi_2)^T$ which, in an orthogonal electromagnetic field with scalar potential $\varphi(\vec{\rho})$ and vector potential $\vec{A}(\vec{\rho})$, is determined by the Dirac-like Hamiltonian

$$\hat{H} = v_F \vec{\sigma} \cdot (-i\hbar\nabla_2 + e\vec{A}) - e\varphi I_2, \tag{1}$$

where $v_F \approx 10^6$ ms$^{-1}$ is the Fermi velocity, $\nabla_2 = \vec{x}\partial/\partial x + \vec{y}\partial/\partial y$, $I_2$ is the $2\times 2$ unit matrix, and $\vec{\sigma} = \sigma_x \vec{x} + \tau\sigma_y \vec{y}$ with $\sigma_{x,y}$ the first two Pauli matrixes. Because of this $\hat{H}$, $\Psi$ is regarded as describing a virtual massless Dirac fermion.

For a graphene electron described by $\Psi$, its velocity operator and OAM operator with respect to a point $\vec{\rho}_0$ are [23]

$$\hat{\vec{v}} = v_F \vec{\sigma}, \quad \hat{\vec{l}}_e = (\vec{\rho} - \vec{\rho}_0) \times m_e \hat{\vec{v}}, \tag{2}$$

with $m_e$ the mass of an electron. Electronic states are usually not eigen-states of $\hat{\vec{l}}_e$. However, for a localized and normalized state $\Psi$, the expectation value $\vec{l}_e = \int_\infty \Psi^+ \hat{\vec{l}}_e \Psi d^2\vec{\rho}$ plays the real role, like the expectation value of $\hat{\vec{v}}$ in graphene's quantum Hall effect [24]. For a graphene electron, the usual relation between the OMM and the OAM still holds. In fact, the state $\Psi$ has the current density $\vec{j} = \Psi^+ \hat{\vec{v}} \Psi$

[2,23]. Therefore the OMM is [25]

$$\vec{\mu} = \frac{1}{2}\int_{\infty}(\vec{\rho}-\vec{\rho}_0)\times(-e\vec{j})d^2\vec{\rho} = -\frac{e}{2m_e}\vec{l}_e. \tag{3}$$

Graphene's electronic states can be calculated for $\tau = +1$ only, since OAM is the same for $\tau = \pm 1$. Suppose

$$L_B = \sqrt{\frac{\hbar}{eB}}, \quad \varepsilon_B = v_F\sqrt{\hbar eB} \tag{4}$$

respectively denote magnetic length and energy quantum. One chooses an arbitrary point $(x_0, y_0)$. By adopting symmetric potential gauge $\vec{A} = (B/2)[-(y-y_0)\vec{x}+(x-x_0)\vec{y}]$, $\varphi = 0$ and using polar coordinates $x = x_0 + \rho\cos\theta$, $y = y_0 + \rho\sin\theta$ one has the eigen-energies (LLs) and eigen-states of $\hat{H}$

$$\varepsilon = \pm\sqrt{2n}\varepsilon_B, \tag{5}$$

$$\begin{cases} \psi_1 = \pm\sqrt{\frac{2}{n}}CL_B e^{i(m-1)\theta}e^{-\rho^2/4L_B^2}\rho^{|m|-1} \\ \quad \times [nL(n_\rho,|m|,\rho^2/2L_B^2)-(n_\rho+|m|)L(n_\rho-1,|m|,\rho^2/2L_B^2)], \\ \psi_2 = iCe^{im\theta}e^{-\rho^2/4L_B^2}\rho^{|m|}L(n_\rho,|m|,\rho^2/2L_B^2) \end{cases} \tag{6}$$

where $n_\rho = 0,1,2,\cdots$, $m = 0,\pm 1,\pm 2,\cdots$, $n = n_\rho + (|m|+m)/2 = 0,1,2,\cdots$, $L(n,m,x) = \sum_{k=0}^{n}(n+m)![k!(n-k)!(m+k)!]^{-1}(-x)^k$ is a generalized Laguerre polynomial with $L(-1,m,x) = 0$, and the normalization constant $C = L_B^{-|m|-1}\sqrt{2^{-|m|-2}n_\rho!/\pi(n_\rho+|m|)!}$. States (6) with $m \geq 0$ were presented in Ref. 22.

An LL (5) is degenerate since it is independent of $m$ and $(x_0, y_0)$. As a result real electronic states could be superposition of states (6) with the same $n$. A possible case can be obtained according to graphene's electronic distribution. Electronic density of each state (6) $|\psi_1|^2 + |\psi_2|^2$ roughly occupies a circular area with radius $\sqrt{2}L_B$ and center $(x_0, y_0)$. For a finite (large) graphene with an area $S$, with spin degeneracy and valley degeneracy not included the number of states $N$ for each LL with a fixed $n$ is that of the magnetic-flux quantum $h/e$ which has the area $h/eB = 2\pi L_B^2$ [26], that is,

$$N = \frac{S}{2\pi L_B^2}. \tag{7}$$

Since $2\pi L_B^2$ is also the area of a state (6), it is supposed that each real state is the superposition of states (6) with the same $n$ and $(x_0, y_0)$ but different $m$. For each LL the $N$ centers of states are regarded as uniformly dispersed in the graphene plane so that the circular areas all together fully cover the plane. This distribution lowers the energy of Coulomb interaction between electrons.

The expectation values of the OAM operator and corresponding OMM for a state (6) are calculated out to be

$$\vec{l}_e = \pm\sqrt{2n}\,m_e v_F L_B \vec{z}, \quad \vec{\mu} = \mp\sqrt{\frac{n}{2}}\,e v_F L_B \vec{z}. \tag{8}$$

Results for the superposed states are also (8), as long as $n$ and $(x_0, y_0)$ are fixed. Therefore states (6) with $\varepsilon < 0$ have paramagnetic moment and those with $\varepsilon > 0$ have diamagnetic one. The OMM (8) is much larger than Bohr magneton $\mu_B = \hbar e/2m_e$, as is found in carbon nanotubes [27]. For instance, for $n=1$ and $B=10\,\text{T}$, one has $l_e \approx 50\hbar$ and $\mu \approx 50\mu_B$. Graphene's magnetism is thus mainly determined by its electrons' OMM. As a result, carrier-undoped graphene could have large para-magnetization, since at temperature $T=0\,\text{K}$ and for Fermi energy $\varepsilon_F = 0$, states (6) with $\varepsilon < 0$ are occupied and those with $\varepsilon > 0$ are empty. Nevertheless, graphene's magnetization cannot be calculated by summing the OMM in Eq. (8), because LLs (5) and wave-functions (6) may not well represent deep states whose energy is far below zero. Instead one can calculate the magnetization variation with carrier-doping. Graphene's magnetization variation can be explains as follows: When the graphene is increasingly doped with electrons, more and more states with $\varepsilon > 0$ are occupied, and this brings about more and more electrons with diamagnetic moment; when the graphene is increasingly doped with holes, more and more states with $\varepsilon < 0$ are empty, and this brings about less and less electrons with paramagnetic moment. In both cases, the magnetization is decreased with the increasing carrier-doping.

Suppose the magnetization for $T=0\,\text{K}$ and $\varepsilon_F = 0$ is $M_0\vec{z}$ and that for $T>0\,\text{K}$ and $\varepsilon_F \neq 0$ is $M\vec{z}$. With spin degeneracy and valley degeneracy included, both the occupied $n$th LL (doped with electrons) and the empty $-n$th LL (doped with holes) contribute to the magnetization by the dia-magnetization $-4N\sqrt{n/2}\,v_F e L_B \vec{z}/S = (-\sqrt{2n}\,v_F e/\pi L_B)\vec{z}$. The magnetization variation $\Delta M \vec{z} = (M - M_0)\vec{z}$ can be calculated by

$$\Delta M = -\frac{e v_F}{\pi L_B}\sum_{n=1}^{+\infty}\sqrt{2n}\left[1 - f(-\sqrt{2n}\,\varepsilon_B) + f(\sqrt{2n}\,\varepsilon_B)\right], \tag{9}$$

where $f(\varepsilon)=1/\{1+\exp[(\varepsilon-\varepsilon_F)/k_B T]\}$ is the Fermi-Dirac distribution function with $k_B$ the Boltzmann constant. Usually the carrier-doping involves only LLs with small $n$, and terms in Eq. (9) with very large $n$ actually do not contribute to the result.

In general the $\Delta M \sim \varepsilon_F$ curve for fixed $B$ and $T$ resembles a parabola. For $\varepsilon_{B,F} \ll k_B T$, by calculating Eq. (9) as an integral one obtains

$$\Delta M = -\frac{2e}{h\varepsilon_B^2}[3\zeta(3)(k_B T)^3 + 2\ln 2 \times k_B T \varepsilon_F^2], \tag{10}$$

with $\zeta(3)=\sum_{k=1}^{+\infty} k^{-3} \approx 1.202$, indicating $|\Delta M| \propto T^3$ for $\varepsilon_F = 0$. For a large magnetic field, the $\Delta M \sim \varepsilon_F$ curve presents plateaus like those in quantum Hall effect [28]. Suppose LLs up to $n$ are filled with electrons or LLs above from $-n$ are filled with holes. In both cases the corresponding $\Delta M$ plateau is

$$\Delta M = -\frac{ev_F}{\pi L_B}\sum_{k=1}^{n}\sqrt{2k}. \tag{11}$$

$\Delta M \sim \varepsilon_F$ curves are presented in Fig. 1. For fixed $T$ and $\varepsilon_F$ the $\Delta M \sim B$ curve demonstrates de Haas-van Alphen oscillation. $\Delta M \sim B$ curves are presented in Fig. 2.

One may also adopt non-symmetric potential gauge $\vec{A}=Bx\vec{y}$, $\varphi=0$ and obtains the same LLs (5) and another set of eigen-states of $\hat{H}$ [21]

$$\begin{pmatrix}\psi_1 \\ \psi_2\end{pmatrix} = \begin{pmatrix}\pm\sqrt{2n}H_{n-1}(\xi) \\ iH_n(\xi)\end{pmatrix} e^{-\xi^2/2+ik_y y}, \tag{12}$$

where $n=0,1,2,\cdots$, $k_y$ is the wave vector, $\xi = L_B^{-1}(x+L_B^2 k_y)$, and $H_n(\xi)=(-1)^n e^{\xi^2} d^n e^{-\xi^2}/d\xi^n$ is the $n$th-order Hermite polynomial with $H_{-1}(\xi)=0$. States (12) are extended in $y$-direction. In general, localized wave-packet states are formed from degenerate states with the same $n$ and similar $k_y$. The OAM can be calculated from a wave-packet's central state $\Psi$ by

$$\vec{l}_e = \lim_{\substack{a\to+\infty \\ b\to+\infty}} \frac{\int_{y_0-b}^{y_0+b}dy\int_{x_0-a}^{x_0+a}\Psi^+\hat{\vec{l}}_e\Psi dx}{\int_{y_0-b}^{y_0+b}dy\int_{x_0-a}^{x_0+a}\Psi^+\Psi dx}, \tag{13}$$

where $(x_0, y_0)$ is the wave-packet center. One obtains

$$\vec{l}_e = \pm\sqrt{\frac{n}{2}}m_e v_F L_B \vec{z}, \quad \vec{\mu} = \mp\sqrt{\frac{n}{8}}ev_F L_B \vec{z}. \tag{14}$$

The result is the same for the potential gauge $\vec{A} = -By\vec{x}$, $\varphi = 0$. Both OAM and OMM of states (12) are different from those of states (6) by a factor and this is the result of the LL degeneracy, which will be further discussed in Section 4.

The degeneracy of LLs can be eliminated by applying a small parallel electric field $\vec{E}$. By using rotated coordinates $x' = x\cos\alpha + y\sin\alpha$, $y' = -x\sin\alpha + y\cos\alpha$ with $\alpha$ the angle between $\vec{E}$, $\vec{x}$ and adopting potential gauge $\vec{A} = Bx'\vec{y}'$, $\varphi = -Ex'$, one has the eigen-energies and eigen-states of $\hat{H}$ [26,29]

$$\varepsilon = \pm(1-\beta^2)^{3/4}\sqrt{2n}\,\varepsilon_B - \beta v_F \hbar k_y, \tag{15}$$

$$\begin{pmatrix} \psi_1 \\ \psi_2 \end{pmatrix} = \begin{pmatrix} e^{-i\alpha}[\pm(\sqrt{1-\beta^2}+1)\sqrt{2n}H_{n-1}(\eta) - \beta H_n(\eta)] \\ i[\mp\beta\sqrt{2n}H_{n-1}(\eta) + (\sqrt{1-\beta^2}+1)H_n(\eta)] \end{pmatrix} e^{-\eta^2/2 + ik_y y'}, \tag{16}$$

where $n = 0, 1, 2, \cdots$, $\beta = E/v_F B < 1$, and $\eta = (1-\beta^2)^{1/4} L_B^{-1}[x' + L_B^2 k_y \pm \beta(1-\beta^2)^{-1/4}\sqrt{2n}\,L_B]$. States (16) are non-degenerate and are the only possible states of graphene electrons in the electromagnetic field. The OAM and OMM of wave packets of states (16) are calculated out to be

$$\vec{l}_e = \pm\frac{\sqrt{2n}\,m_e v_F L_B}{2\sqrt{1-\beta^2}}\vec{z}, \quad \vec{\mu} = \mp\frac{\sqrt{2n}\,ev_F L_B}{4\sqrt{1-\beta^2}}\vec{z}. \tag{17}$$

For the $n$ th LL of a finite graphene with dimensions $L_x$, $L_y$ respectively in $x$- and $y$-directions, $k_y$ is determined by [24]

$$k_y = \frac{2\pi n_y}{L_y}, \quad -\frac{L_x L_y}{4\pi L_B^2} \mp \frac{\beta\sqrt{2n}\,L_y}{2\pi(1-\beta^2)^{1/4} L_B} \leq n_y \leq \frac{L_x L_y}{4\pi L_B^2} \mp \frac{\beta\sqrt{2n}\,L_y}{2\pi(1-\beta^2)^{1/4} L_B}, \tag{18}$$

with $n_y$ an integer. The magnetization variation is

$$\Delta M = -\frac{ev_F L_B}{\sqrt{1-\beta^2}\,L_x L_y}\left\{\sum_{n,n_y,\varepsilon<0}\sqrt{2n}[1-f(\varepsilon)] + \sum_{n,n_y,\varepsilon>0}\sqrt{2n}\,f(\varepsilon)\right\}, \tag{19}$$

where $n = 0, 1, 2, \cdots$, $n_y$ is given by Eq. (18) and $\varepsilon$ by Eq. (15). The regulation of graphene's magnetization variation by an electric field is presented in Fig. 3 and the effect is similar to that of temperature.

## 3. Origin of the orbital angular momentum and Dirac fermions

Different from graphene electrons, the Dirac fermions have the OAM operator [23, 30]

$$\hat{\vec{l}}_D = (\vec{\rho} - \vec{\rho}_0) \times I_2(-i\hbar\nabla_2 + e\vec{A}). \tag{20}$$

States (6), (12), and (16) are not eigen-states of $\hat{\vec{l}}_D$ either, and their expectation values are respectively calculated out to be

$$\vec{l}_D = 2n\hbar\vec{z}, \quad \vec{l}_D = n\hbar\vec{z}, \quad \vec{l}_D = \left[\frac{n\hbar}{\sqrt{1-\beta^2}} - \frac{\sqrt{2n}\beta\hbar k_y L_B}{2(1-\beta^2)^{1/4}}\right]\vec{z}. \tag{21}$$

These values are much smaller than those of $\vec{l}_e$. In terms of OAM, a graphene electron is different from the Dirac fermion, although they are described by the same wave function $\Psi$. The difference originates from graphene's crystal structure which is absent for the Dirac fermion. Originally, a graphene electron is described by the full wave function $\psi(\vec{r},t) = \psi_1(\vec{\rho},t)\psi_A(\vec{r}) + \psi_2(\vec{\rho},t)\psi_B(\vec{r})$, where $\psi_{A,B} = \sqrt{\Omega}e^{\pm i\pi/12}\sum_{A\text{ or }B} e^{i\vec{k}_F\cdot\vec{r}_{A,B}}\phi_{A,B}(\vec{r})$ are Bloch functions at $\vec{k}_F$ respectively for type-$A$ and type-$B$ atoms, with $\Omega$ the area of a unit cell and $\phi_{A,B}$ carbon $2p_z$ orbitals at $\vec{r}_{A,B}$. The electron's OAM operator which acts on $\psi$ is the standard one $\hat{\vec{l}} = (\vec{r}-\vec{\rho}_0)\times(-i\hbar\nabla + e\vec{A})$ with $\nabla = \nabla_2 + \vec{z}\partial/\partial z$. Suppose $g(\vec{\rho},t)$ is a function that varies gently at atomic scale. Due to graphene's lattice structure, $\psi_{A,B}$ have the following properties (each equation represents two formulas respectively for the two subscripts) [23]: $\int_\infty g(\vec{\rho},t)\psi_{A,B}^*\psi_{A,B}d^3\vec{r} = \int_\infty g(\vec{\rho},t)d^2\vec{\rho}$, $\int_\infty g(\vec{\rho},t)\psi_{A,B}^*\psi_{B,A}d^3\vec{r} = 0$, $\int_\infty g(\vec{\rho},t)\psi_{A,B}^*\nabla\psi_{A,B}d^3\vec{r} = 0$, and $\int_\infty g(\vec{\rho},t)\psi_{A,B}^*(-i\hbar)\nabla\psi_{B,A}d^3\vec{r} = m_e v_F(\vec{x}\mp\tau i\vec{y})\int_\infty g(\vec{\rho},t)d^2\vec{\rho}$. By applying these properties to the expectation value $\vec{l} = \int_\infty \psi^*\hat{\vec{l}}\psi d^3\vec{r}$ one has

$$\vec{l} = \int_\infty \Psi^+\hat{\vec{l}}_e\Psi d^2\vec{\rho} + \int_\infty \Psi^+\hat{\vec{l}}_D\Psi d^2\vec{\rho}. \tag{22}$$

Because the second integral in Eq. (22) is small, one obtains Eq. (2). More specifically, the vector hopping interaction $\int_\infty \phi_{A,B}\nabla\phi_{B,A}d^3\vec{r}$ between $2p_z$ orbitals of two neighboring atoms is the building element of the OAM of graphene electrons. The three directions of $\int_\infty \phi_{A,B}\nabla\phi_{B,A}d^3\vec{r}$ along three $sp^2$ hybridized carbon orbitals bring about the vector Pauli matrix $\vec{\sigma}$ by $-i\hbar\sum_{B\text{ or }A} e^{i\vec{k}_F\cdot(\vec{\rho}_{B,A}-\vec{\rho}_{A,B})}\int_\infty \phi_{A,B}\nabla\phi_{B,A}d^3\vec{r} = m_e v_F e^{\pm\pi i/6}(\vec{x}\mp\tau i\vec{y})$ [23,31], and its intensity determines $v_F$ by $m_e v_F = (3\hbar/2)\left|\int_\infty \phi_{A,B}\nabla\phi_{B,A}d^3\vec{r}\right|$ [4]. An OAM is in fact the superposition of $(\vec{\rho}_{A,B}-\vec{\rho}_0)\times e^{i\vec{k}_F\cdot(\vec{\rho}_{B,A}-\vec{\rho}_{A,B})}\int_\infty \phi_{A,B}\nabla\phi_{B,A}d^3\vec{r}$ for all atoms, with the superposition coefficients given by $\Psi$. Therefore the OAM reflects the electronic motion at the atomic scale within the scope of state $\Psi$. It is the internal structure of the wave packet, not its integral motion that determines the OMM. In fact, states

(6) have zero expectation value of the velocity operator and the wave packets are stationary, but their OAM is nonzero. Although wave packets of states (16) have the velocity $-\beta v_F \vec{y}'$ [24], their OMM is not generated by this velocity, rather it is the result of the electronic motion inside the wave packets. Finally, as normalized electronic states have the dimension of $L_B$, OAM has the order of $m_e v_F L_B$.

Because the velocity operator of the Dirac fermion is the same as that of the graphene electron [23], the Dirac fermion's OMM is also determined according to Eq. (3) by $\vec{l}_e$, not by $\vec{l}_D$. One may introduce the OMM operator for both graphene electrons and Dirac fermions

$$\hat{\vec{\mu}} = -\frac{e}{2}(\vec{\rho} - \vec{\rho}_0) \times \hat{\vec{v}}, \tag{23}$$

but its relation to $\hat{\vec{l}}_D$ cannot be established. The significance of $\hat{\vec{l}}_D$ is demonstrated as follows: According to Eq. (22), the more exact OAM operator of graphene electrons is

$$\hat{\vec{l}}'_e = \hat{\vec{l}}_e + \hat{\vec{l}}_D \tag{24}$$

with $\hat{\vec{l}}_e$ its main part. If the OAM is approximated by $\hat{\vec{l}}_e$, then $\hat{\vec{l}}_D$ can be regarded as a new spin of a graphene electron other than its true spin and pseudo-spin. States (6) are eigen-states of the total-angular-momentum operator of field-free Dirac fermions

$$\hat{\vec{l}}'_D = (\vec{\rho} - \vec{\rho}_0) \times (-i\hbar)I_2\nabla_2 + \frac{\hbar}{2}\sigma_z \vec{z} \tag{25}$$

with $\sigma_z$ the third Pauli matrix, and corresponding eigen-values are $(m-1/2)\hbar\vec{z}$ [22]. This reflects the isotropy of the graphene plane to Dirac fermions, since the graphene lattice is effective only for the electrons. The effect of the magnetic field is to realize eigen-states of the field-free operator $\hat{\vec{l}}'_D$, not those of the operator $\hat{\vec{l}}_D$ with the field, and graphene's magnetism is determined by the expectation values of $\hat{\vec{l}}_e$, neither those of $\hat{\vec{l}}_D$ nor those of $\hat{\vec{l}}'_D$.

## 4. Further discussion

The difference between OMM (8) and OMM (14) demonstrates the effect of LL degeneracy. This may be related to some fundamental problems of quantum mechanics. Usually to eliminate degeneracy, additional mechanical quantities are introduced so that the states become the common eigen-states of a complete set of operators. The additional operators for graphene electrons in a magnetic field remain an interesting problem. It is noted that states (6) are also eigen-states of operator (25) and states (12) are not. Graphene's real electronic states may also be determined by environmental conditions or by the establishing process of the magnetic field. These unsolved problems call for further experimental investigations. If stochastic conditions determine the real electronic states, randomness in magnetization will be observed. If definite magnetization is observed, systematic reasons exist in determining the electronic states. In all cases, de Haas-van Alphen oscillation or plateaus in the magnetization variation are expected to be observed. Besides, magnetization becomes definite when an electric field is applied. As a result, the abrupt variation in graphene's magnetization may be observed.

Graphene was observed to be diamagnetic experimentally [20] and its dia-susceptibility was derived theoretically according to the LLs [1,3,4]. From wave functions (6) or (12), however, different result may be obtained because according to OMM (8) or (14), undoped graphene's occupied states with $\varepsilon < 0$ contribute a paramagnetic moment. It is interesting to compare this with the diamagnetism of two-dimensional free electrons (2DFEs). 2DFEs in a perpendicular magnetic field have eigen-energies $(n+1/2)\hbar e B m_e^{-1}$ with $n = 0,1,2,\cdots$ and eigen-states $\sqrt{2}Ce^{im\theta}e^{-\rho^2/4L_B^2}\rho^{|m|}L(n_\rho,|m|,\rho^2/2L_B^2)$, and their OAM and OMM are $(2n+1)\hbar\vec{z}$ and $-(n+1/2)\hbar e m_e^{-1}\vec{z}$ respectively. All the eigen-energies are positive and the eigen-states contribute a diamagnetic moment, which is in accordance with the diamagnetism of free-electron gas. For undoped graphene, it is unknown whether it is the superposition of states (6) or (12) that results in the observed diamagnetism, or the diamagnetism should be attributed to the deep states that cannot be presented by Eq. (6) or (12). Besides, the single-electron states could be modified if more exact electron-electron interaction is considered. On the other hand, results in this work are valid for perfect infinite graphene, since the electronic states depend on the translational symmetry of graphene lattice. They may be approximately correct for a finite graphene with a dimension much greater than $L_B$, where the boundary conditions can only affect a few electronic states. For a finite graphene with a dimension comparable to $L_B$, the broken translational symmetry of graphene lattice plays an important role and the electronic states are no longer given by Eqs, (6), (12), and (16). The weaker the magnetic field, the more evident the size effect is, since $L_B \propto B^{-1/2}$. As a result, the finite size of graphene may also play a role in the experimentally observed diamagnetism. In fact, even the experiments of graphite were carried out for crystallites, from the early graphite powder [32] to later highly oriented pyrolytic graphite (HOPG) [33,34]. Efforts were mainly focused on the orientation of the crystallites.

## 5. Conclusion

In conclusion, electronic wave functions of graphene in a magnetic field may provide more diverse results about its magnetism than LLs. Along with further experimental investigations, these results may provide new insight into graphene's electronic structure and the effect of electron-electron interaction. Results of this work may also help to understand the magnetism of perfect graphite.

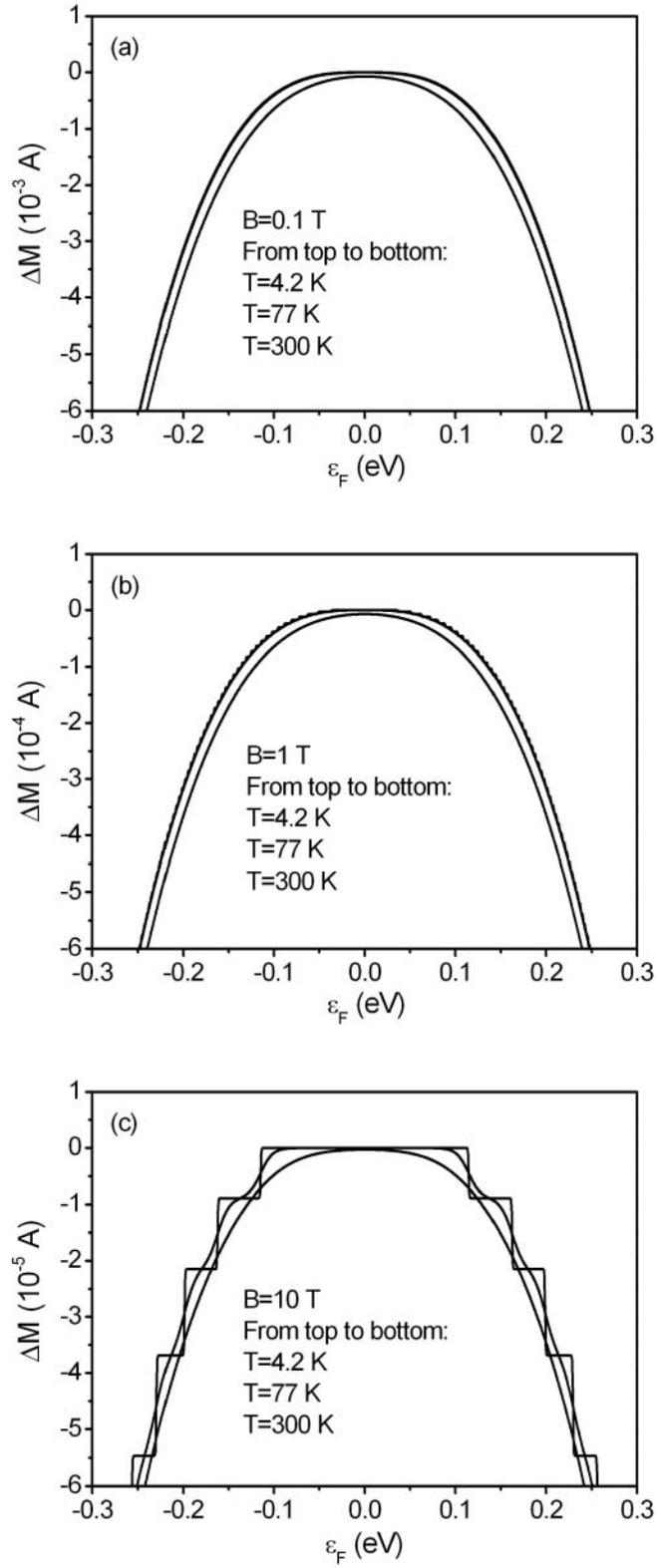

Fig. 1. Graphene's magnetization variation as a function of Fermi energy for different magnetic field and at different temperature, calculated according to Eq. (9).

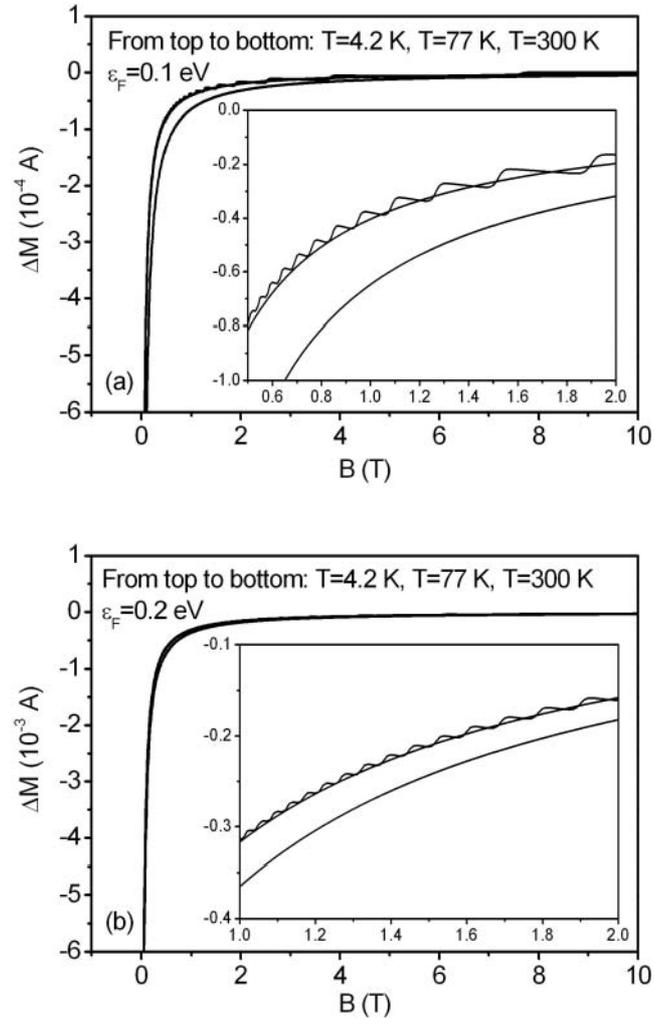

Fig. 2. Graphene's magnetization variation as a function of magnetic field for different Fermi energy and at different temperature, calculated according to Eq. (9).

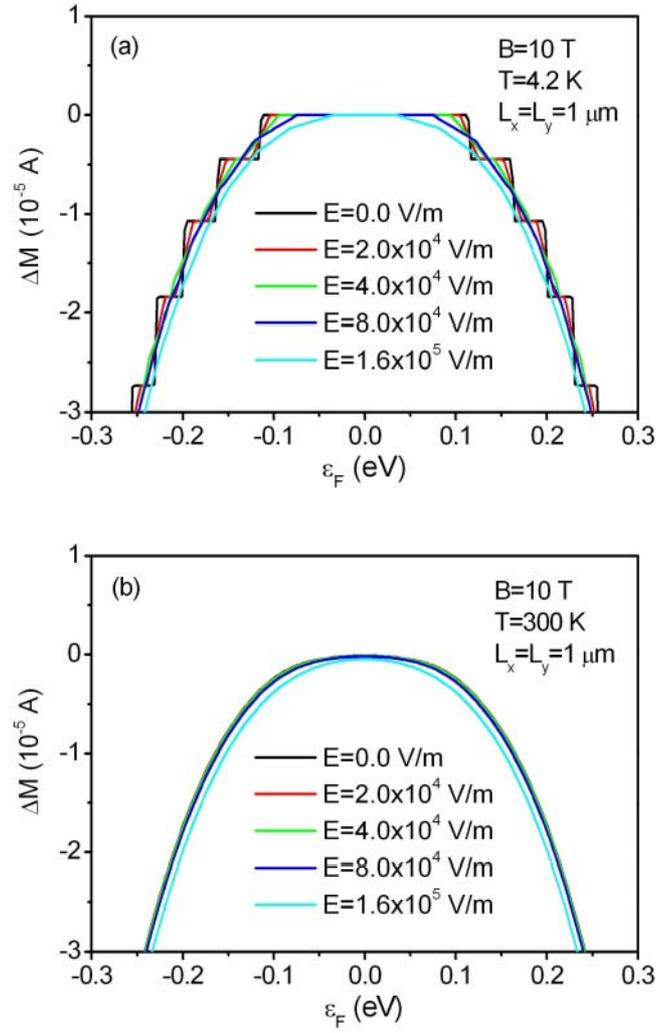

Fig. 3. Graphene's magnetization variation as a function of Fermi energy in different electric field and at different temperature, calculated according to Eq. (19).